\documentclass[twocolumn,prl,aps]{revtex4}
\usepackage{amsmath}
\usepackage{amssymb}
\usepackage{amsfonts}
\usepackage{graphicx}

\begin{document}
\title
{Azbel'-Kaner-Like Cyclotron Resonance in a Two-Dimensional Electron System}

\author{I.~V.~Andreev\footnote{Both authors contributed equally to this work}, V.~M.~Muravev\footnotemark[1], V.~N.~Belyanin, I.~V.~Kukushkin}
\affiliation{Institute of Solid State Physics, RAS, Chernogolovka, 142432 Russia}
\date{\today}
\date{\today}

\begin{abstract}
Resonant microwave absorption of a two-dimensional electron system in an AlGaAs/GaAs heterostructure excited by a near-field technique was investigated. Along with collective magnetoplasmon modes, we observed resonance that precisely follows the cyclotron resonance (CR) position and revealed no signs of collective plasma depolarization shift. We show that the discovered CR mode is absent in the Faraday geometry, and is localized at the edge of the exciting metal electrode. Such behavior points in favor of the single-particle Azbel'-Kaner nature of the discovered resonance.
\end{abstract}

\maketitle

Cyclotron resonance (CR) is one of the most significant phenomena in condensed matter physics. CR spectroscopy is widely used to study energy band structure and charge carrier properties in semiconductors and metals. However, CR excitation configurations in semiconductors and metals are quite different. In semiconductors, CR is observed in Faraday geometry~\cite{Dresselhaus}, when electromagnetic radiation propagates along the direction of magnetic field $B$. In this geometry, CR absorption occurs only for one sense of light circular polarization with respect to the $B$ field. In this case, the CR is observed at frequencies where the electromagnetic radiation penetrates the sample, that is, at frequencies above the plasma frequency ($\omega > \omega_p$)~\cite{Dresselhaus}. Therefore, in semiconductors, electrons move in an almost homogeneous electromagnetic field.  By contrast, in metals, the CR is observed in Voigt geometry at frequencies $\omega < \omega_p$, where, because of high electron screening, the electromagnetic radiation decays in a thin skin layer near the sample surface. The amplitude of the electromagnetic field has a large inhomogeneity over the cyclotron orbit of the carriers. The incident oscillating electric field in the skin layer pushes electrons, which circulate the CR orbit in the bulk, and afterwards synchronously return back to the skin layer. The resonant phenomenon observed under such a condition is called Azbel'-Kaner cyclotron resonance (AK-CR)~\cite{Azbel, Fawcett}.  

Passing electromagnetic radiation through a two-dimensional electron system (2DES), placed in a magnetic field, one would observe the collective magnetoplasma cyclotron resonance, which is shifted to a higher frequency because of a depolarization effect~\cite{Allen:83, Heitmann:88, Studenikin:05, Shaner:13, Lusakowski:15, Zudov:16, Scalari:17}. This is very similar to the CR observed in semiconductors. The frequency of the collective cyclotron magnetoplasma mode is described by the equation~\cite{Chaplik:72}
\begin{equation}
\omega^2=\omega_p^2 + \omega_c^2,
\label{f1}
\end{equation} 
where $\omega_c=eB/m^{\ast}$ is the cyclotron frequency ($m^{\ast}=0.067m_0$ is an effective electron mass in GaAs), and $\omega_p$ is the dimensional plasmon frequency. All prior experiments were conducted in the quasioptical Faraday setup, when the cyclotron magnetoplasma mode is excited by the electromagnetic radiation passing through the 2DES from free space along the $B$-field direction. 

The Voigt setup is not well defined for the 2DES. However, near-field excitation could mimic the experiment in Voigt geometry. In the present research, we developed a near-field 2D plasma excitation technique. Using this technique, we observed a new single-particle cyclotron resonance along with collective magnetoplasma excitations. It turned out that the resonance is absent in the Faraday geometry. It is shown that the single-particle CR is excited in the region of a locally nonuniform high-frequency electromagnetic field near the metallic gate edge. Therefore, the discovered resonance has some similarity with Azbel'-Kaner cyclotron resonance observed in metals.  

The experiments were performed on a $25$-nm-wide GaAs/AlGaAs single quantum well, which was located $480$~nm underneath the crystal surface. A schematic of the sample geometry is depicted in Fig.~1. The sample mesa had the shape of a 2DES disc of diameter $D=1$~mm, which was grounded by a perimetric ohmic contact. To make a near-field axisymmetric 2D plasma excitation, a disc-shaped gate of considerably smaller diameter ($d=0.1$~mm) was fabricated at the center of the mesa (quasi-Corbino geometry)~\cite{Muravev:17}. Microwave radiation with frequencies from $1$ to $40$~GHz was guided into the cryostat with a coaxial cable, and transferred to the excitation gate via an impedance-matched coplanar waveguide transmission line. Electron density in 2DES was $n_s=1.0\times10^{11}~\text{cm}^{-2}$ with mobility $\mu=5\times10^6~\text{cm}^2/\text{V$\cdot$s}$ at temperature $T=1.5$~K. To detect 2D electron plasma excitations, we employed a non-invasive optical technique of microwave absorption detection~\cite{Ashkinadze, Kukushkin:02}. This technique is based on the high sensitivity of the 2DES recombination luminescence spectrum on the heating caused by the absorption of incident microwave radiation. In this technique, 2DES luminescence spectra with and without microwave radiation were collected using a spectrometer with a liquid nitrogen-cooled CCD camera. The resulting differential spectrum reveals microwave-caused 2DES temperature heating. Therefore, the integral of the absolute value of the differential spectrum could be used as a measure of microwave absorption by the 2DES. In our experiments, we used a $780$~nm stabilized semiconductor laser with $4$~mW net output power to excite 2D luminescence and a single-grating spectrometer with $0.25$~meV spectral resolution to analyze the luminescence spectra. All the experiments were carried out in the liquid helium with a superconducting magnet at temperature $T=1.5$~K. The magnetic field was directed perpendicular to the sample surface and swept in the range $0 - 0.5$~T.

\begin{figure}[!t]
\center
\includegraphics[width=0.47 \textwidth]{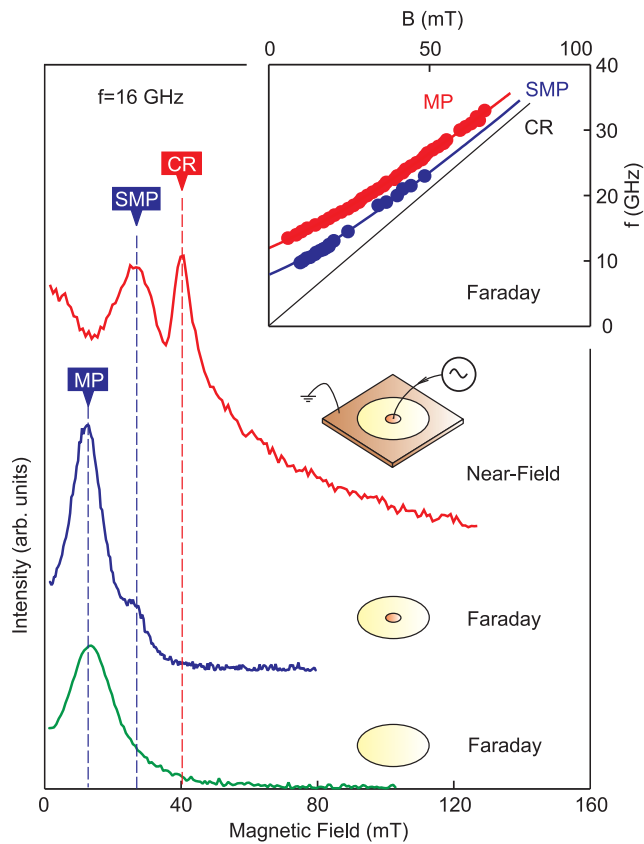}
\caption{Dependence of microwave absorption versus magnetic field for different sample geometries and excitation techniques. Sample schematics are drawn next to the corresponding curves. The excitation frequency is $f=16$~GHz for all curves. Fundamental magnetoplasmon (MP), screened magnetoplasmon (SMP) and cyclotron resonance (CR) magnetic field positions at $f=16$~GHz are marked by labels and vertical dashed lines. MP and SMP modes are excited in both Faraday and near-field geometries, whereas CR is excited only in near-field geometry. The inset plots magnetodispersions of MP (red circles) and SMP (blue circles) modes for a disc sample of $D=1$~mm outer diameter and $d=0.1$~mm concentric central gate. Solid curves correspond to the theoretical dependencies. CR magnetodispersion is also indicated.}
\label{1}
\end{figure}

In our experiments, we compared spectra of microwave absorption in three different 2DES samples, excited in Faraday and near-field geometries. Sample schematics are shown in Fig.~\ref{1}: (1) a single 2DES disc of diameter $D=1$~mm, excited via rectangular waveguide in Faraday geometry; (2) a 2D disc of identical geometry with a concentric $d=0.1$~mm gate, excited in the same configuration; and (3) a $D=1$~mm disc with a perimetric grounded contact and a concentric $d=0.1$~mm gate, excited via the coaxial cable (quasi-Corbino geometry). In Fig.~\ref{1}, we present typical microwave absorption dependencies versus magnetic field measured on these three experimental setups at $16$~GHz. To understand the nature of the plasma waves involved in our investigation, the experiment of Fig.~\ref{1} was
repeated for a large set of microwave frequencies. Resultant magnetodispersions for observed resonances are depicted in the inset of Fig.~\ref{1}.

There is only one resonant peak for a single 2DES, excited in the Faraday geometry (green curve in Fig.~\ref{1}). The magnetodispersion of this peak shows the typical cyclotron magnetoplasmon (MP) mode behavior that has been observed on many occasions in quantum dots and classical discs (red circles in the inset to Fig.~\ref{1})~\cite{Allen:83, Glattli:85, Heitmann:90}. The MP mode approaches the cyclotron resonance (solid line) but never crosses it. The measured spectrum agrees well with the theoretical result for the fundamental MP (solid line in Fig.~\ref{1} inset)~\cite{Allen:83, Shikin}       

\begin{equation}
\omega=\pm \frac{\omega_c}{2} + \sqrt{\omega_p^2+\left(\frac{\omega_c}{2}\right)^2},
\label{f2}
\end{equation}

where $\omega_p$ is the plasmon frequency for $B=0$~T. The frequency $\omega_p$ obeys the 2D-plasmon dispersion~\cite{Stern:67}:

\begin{equation}
\omega_p=\sqrt{\frac{n_s e^2}{2 m^{\ast} \varepsilon \varepsilon_0} q},
\label{f3}
\end{equation}

where $\varepsilon=(1+\varepsilon_{\rm GaAs})/2$ is the effective dielectric permittivity of the surrounding medium and $q=2.4/d$ is the wave vector for the disc geometry~\cite{Kukushkin:03}. From our experiment, we measured zero-field plasma frequency $f_p=\omega_p/{2 \pi} = 13$~GHz, which coincides with the value obtained from Eq.~(\ref{f3}).      
There are two resonances for a 2DES disc with a concentric $d=0.1$~mm top gate, excited in the Faraday geometry (blue curve in Fig.~\ref{1}). The magnetic field position of the fundamental MP mode remains almost unchanged. However, a second MP mode arises at larger values of $B$-field values. The magnetodispersion of these two modes is summarized in the inset of Fig.~\ref{1}. The second mode (blue circles) also follows  the magnetodispersion law predicted by  Eq.~(\ref{f2}), but with a lower plasmon frequency of  $f_p=8$~GHz. This mode corresponds to the screened magnetoplasmon (SMP), localized beneath the gate. The SMP frequency is described by the following formula~\cite{Fetter}: 
\begin{equation}
\omega_p=\alpha'_{11}\sqrt{\frac{2 \, n_s e^2 h}{ m^{\ast} \varepsilon \varepsilon_0}} \frac{1}{d},
\label{f4}
\end{equation}
where $\alpha'_{11} \approx 1.84$ is the first zero of the $J'_1(x)$ Bessel function, and $h=480$~nm is the distance between the 2DES and the top gate. In our case, the theoretically calculated $f_p=\omega_p/2\pi=7.6$~GHz agrees well with the experiment. It is noteworthy to mention that the same experiment was repeated for the same sample with an array of gate discs ($N=19$) evaporated on top of the mesa. In this case, the structure of the magnetoplasmon modes remained the same with no extra resonances.     

\begin{figure}[!t]
\includegraphics[width=0.47 \textwidth]{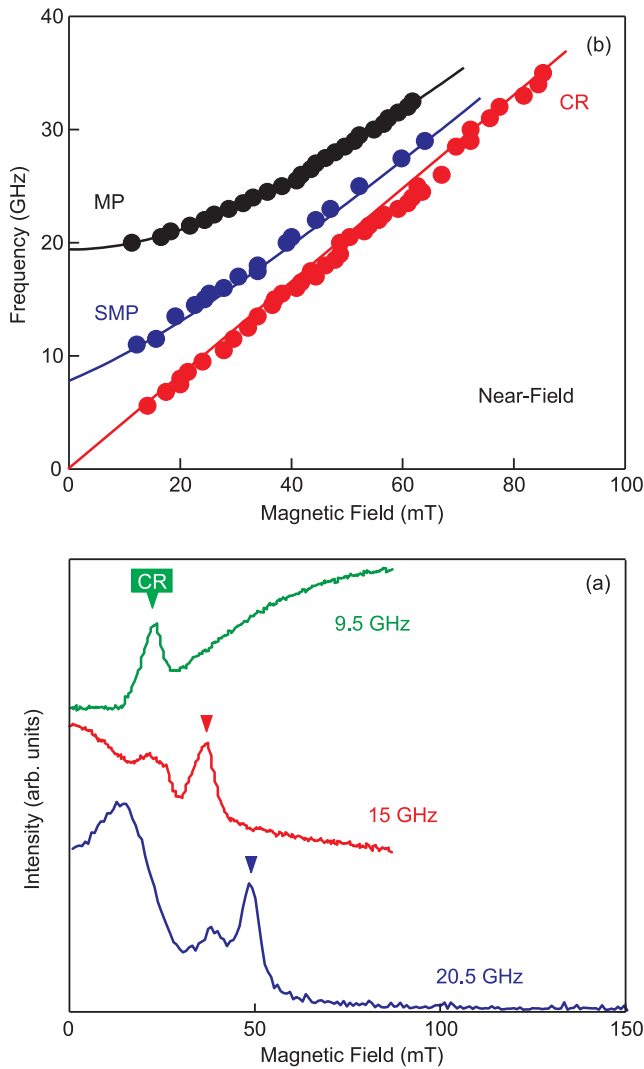}
\caption{(a) Microwave absorption as a function of the magnetic field measured for a series of microwave frequencies ($f=9.5$, $15$ and $20.5$~GHz). The sample had a quasi-Corbino geometry with microwave radiation guided directly to the central top gate. Curves are offset vertically for clarity. (b) The magnetic-field position of the dark axisymmetric magnetoplasmon (MP), screened magnetoplasmon (SMP), and cyclotron (CR) resonances versus excitation frequency.}
\label{2}
\end{figure}

The situation drastically changes when we guide microwave radiation directly to the central top gate. In this near-field excitation setup, SMP mode stays at the same $B$-field position and becomes more pronounced. Unexpectedly, a new resonance emerges at the exact position of cyclotron resonance $B_c= \omega m^{*}/e$. 
The important feature of the discovered CR mode is that it is observed only using the near-field technique, when the central gate is directly wired to the microwave generator via the coaxial cable. The resonance is absent in the Faraday geometry, when the radiation is incident on the sample from free space (Fig.~\ref{1}). The discovered CR resonance is thoroughly studied in the remainder of this paper.   

In Fig.~\ref{2}(a) microwave absorption is plotted for a few different radiation frequencies $f=9.5, 15, 20.5$~GHz. Absorption curves are offset vertically for clarity. The curves reveal from one to three resonant peaks. The origin of these resonances is best identified by plotting their magnetic-field positions versus the incident microwave frequency as in Fig.~\ref{2}(b). The strongest resonance arises at the frequency range $f>19$~GHz. This mode corresponds to excitation of the fundamental dark magnetoplasmon mode (MP), i.e., a standing plasma wave with an antinode at the center and a node at the grounded circumference of the disc. This mode has been well studied elsewhere~\cite{Muravev:17}. The second mode that starts at $f=8$~GHz is the screened magnetoplasmon (SMP), located beneath the central gate.
The third mode, marked by arrows in Fig.~\ref{2}(a), has a number of distinctive properties, and, to the best of our knowledge, has not been observed in prior art. Its magnetodispersion with high accuracy follows the CR position and reveals no signs of a collective depolarization shift. Such a behavior is totally unexpected in a 2DES of finite size, and points in favor of the single-particle nature of the discovered resonance. 

\begin{figure}[!t]
\includegraphics[width=0.47 \textwidth]{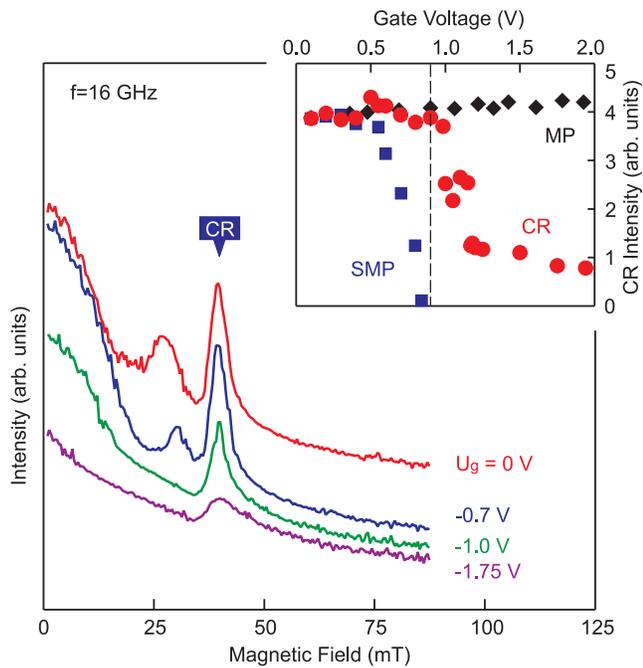}
\caption{Microwave absorption versus magnetic field measured at different values of gate bias ($f=16$~GHz). The CR position $B_c=\omega m^{*}/e=38$~mT is marked by an arrow. Inset: Plots of the normalized resonance amplitude vs. gate bias for dark axisymmetric magnetoplasmon (MP), screened magnetoplasmon (SMP), and cyclotron resonance (CR) modes.}
\label{3}
\end{figure}

The obtained results allow us to suppose that the CR mode is localized in the near-gate region of the 2DES, where excitation takes place. To prove this, we applied a DC bias voltage $U_g$ to the central excitation gate, while keeping the perimetric contact grounded. The bias voltage was negative to deplete the region under the gate. Fig.~\ref{3} shows curves of microwave absorption ($f=16$~GHz) measured at different values of gate bias. Both SMP and CR modes are clearly discernable in the signal. We found that with the depletion of the two-dimensional electrons under the gate, the SMP and CR resonances diminish in a very different way. The SMP mode is highly sensitive to the gate bias and abruptly disappears between $U_g=-0.55$~V and $U_g=-0.9$~V (inset of Fig.~\ref{3}). We note that the gate bias, which corresponds to full depletion of 2DES under the gate is $U_g=-0.9$~V (dashed line in the inset of Fig.~\ref{3}). Such a rapid fadeaway of the SMP mode is no surprise. Indeed, the SMP plasma wave is associated with the oscillation of charges over the whole 2DES area underneath the gate. By contrast, the discovered CR resonance responds more weakly to the gate bias. There is still a distinguishable resonance at the CR position even at $U_g=-2.0$~V, which is well above the full depletion level. This experimental finding confirms that the discovered CR mode is localized at the edge of the exciting gate. Indeed, a pronounced weakening of the CR mode between $U_g=-0.9$~V and $U_g=-1.2$~V corresponds to the full depletion of the near-gate outer region of the 2DES at distances of $0.2$~$\mu$m (see Supplemental Material~\textrm{I}). This distance is comparable to the size of a small region with an extremely strong electric field gradient situated at the circumference of the gate. The size of this inhomogeneity region for our structure geometry is dictated by the distance from the gate to the 2DES, $h=480$~nm~\cite{Mikhailov:06, Mikhailov:11}. This region plays a similar role as the skin layer in the Azbel'-Kaner cyclotron resonance (AK-CR) effect in metals~\cite{Azbel, Fawcett}. Specifically, the gate oscillating electric field in the inhomogeneity region pushes electrons, which circulate the CR orbit in the 2DES ($r_c\approx 1.4$~$\mu$m at the magnetic field of $B=38$~mT), and afterwards, synchronously return back to the region. This mechanism is analogous to what happens in a cyclotron.    

Despite the apparent similarity of the discovered CR mode with the AK-CR effect, there are a number of distinctions. The most significant feature is that unlike in the AK-CR effect, we do not observe resonances at the CR harmonics $B_N=\omega m^{*}/eN$ ($N=2,3 \cdots$). We explain this experimental feature as follows. The necessary condition for CR harmonics observation is strong electric field inhomogeneity over the cyclotron orbit. This condition is easily fulfilled in metals. However, in our experiments, the field variation distance is of the order of $h=480$~nm, which is comparable to $r_c$ and insufficient to detect the CR harmonics. Observation of CR harmonics in 2D plasma is an interesting task and will be a topic for our future research. 

In conclusion, we discovered a new resonant mode, which exactly follows a single-particle cyclotron resonance $\omega_c=eB/m^{\ast}$. The CR mode (1) reveals no signs of a collective depolarization shift, (2) could be excited only using a near-field, but not Faraday, configuration, and (3) is localized in a narrow region near the edge of the exciting gate. All these features make the observed mode similar to the Azbel'-Kaner cyclotron resonance in metals. These experiments pave the way for research in the field of interplay between collective and single-particle phenomena in 2D plasma.    

We thank S.~I.~Gubarev and V.~B.~Shikin for useful discussions and comments. Further, we gratefully acknowledge financial support from the Russian Science Foundation (Grant~No.~14-12-00693).

\end{document}


\title{Supplementary Material for\\ ``Azbel'-Kaner-Like Cyclotron Resonance in a Two-Dimensional Electron System''}

\author{I.~V.~Andreev, V.~M.~Muravev, V.~N.~Belyanin, I.~V.~Kukushkin}
\affiliation{Institute of Solid State Physics, RAS, Chernogolovka, 142432 Russia}

\date{\today}\maketitle

\section{\textrm{I}. 2D electron density distribution at different gate voltages}

To analyze and interpret an effect of gate depletion on the discovered CR mode, we calculated the 2D electron density profile for different values of gate bias $U_g$. A theshold voltage, required to fully deplete 2DES under the gate of infinite lateral dimension is
$$U_0 = -\frac{e n_s d}{\varepsilon \varepsilon_0}.$$

For the structure under study ($d=480$~nm, $n_s=1.0\times10^{11}~\text{cm}^{-2}$ and $\varepsilon = 12.8$), the threshold voltage equals $U_0\approx-0.7$~V.
But, it is intuitively obvious, that at $U_g=U_0$ the electron density in the 2DES under the edge of the gate remains nonzero. One needs further increase in $U_g=U_{\rm depl} > U_0$ to deplete the region of 2DES under the gate edge. An even greater voltage increase, depletes electrons in the region of width $a$ outside the gate edge. Further, we would like to present calculations, which are based on electrostatical approach developed by Larkin and Davies for a semi-infinite half-plane gate~\cite{Larkin}. Although our experiments are carried out on the sample with a disk-shaped gate, we still can rely on the theoretical results, since we are interested only in density distribution of a narrow 2DES region at the gate circumference. The size of the region is much smaller than the gate radius. According to the theoretical prediction, the width $a$ of depleted region outside the gate is~\cite{Larkin}     

$$a=\frac{d}{\pi}(\alpha + \ln\alpha-1),$$

where

$$\alpha = \frac{U_g}{U_0}-1.$$

The whole region under the gate becomes fully depleted whenever $a=0$, which corresponds to $U_{\rm depl} \approx 1.28 \, U_0$ ($U_{\rm depl} \approx -0.9$~V for our samples). If we choose an $x$-axis directed outward the gate perpendicular to the gate edge, and assume that $x=0$ sits at the point exactly under the gate edge (inset to Figure~1S), than electron density distribution could be expressed as follows~\cite{Larkin}
$$\frac{n(x)}{n_s}=\frac{\alpha}{\alpha+\xi_1}-\frac{\alpha}{\alpha+\xi_2},$$
where $\xi_1 < \xi_2$ are roots of the equation $\xi-1-\ln\xi=\pi(x-a)/d$. The 2DES occupies region $x>a$, and $a$ depends strongly on the gate voltage. We calculated values of $a$ for several gate bias voltages $U_g = -0.9, -1.2, -2$~V, which are relevant for our experiments.

\begin{center}
\begin{tabular}{|c|c|}
\hline
\rule{0pt}{3ex}
$U_g$~(V) & $a$~($\mu$m)\\
\hline
\rule{0pt}{3ex}
-0.9 & 0\\
\hline
\rule{0pt}{3ex}
-1.2 & 0.2\\
\hline
\rule{0pt}{3ex}
-2.0 & 0.5\\
\hline
\end{tabular}
\end{center}

\begin{figure}[!t]
\center
\includegraphics[width=0.8\textwidth]{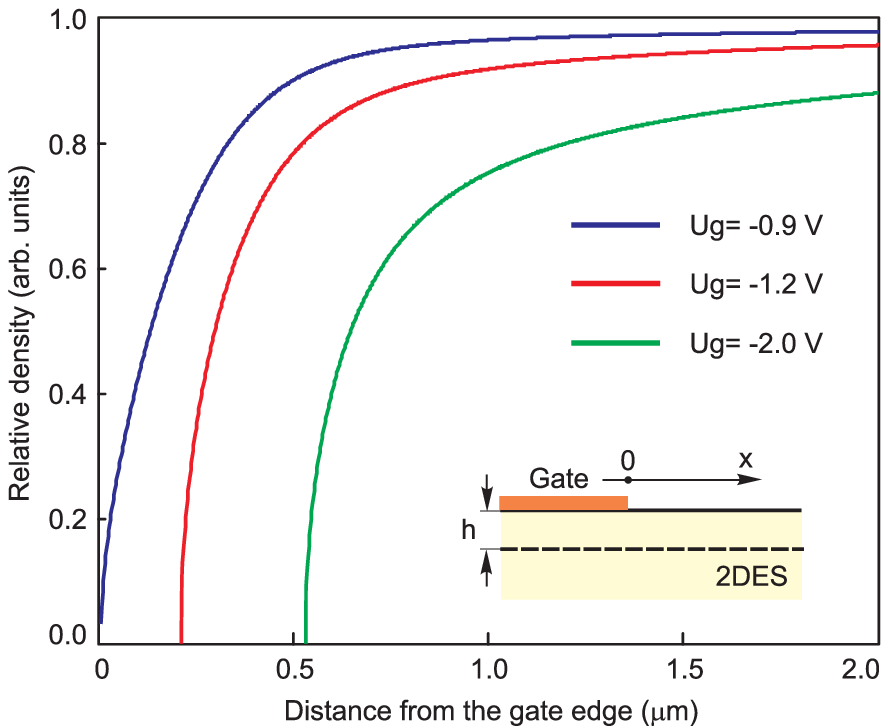}
\caption{Electron density distribution calculated for several gate bias voltages $U_g = -0.9, -1.2, -2$~V. The $x$-axis is directed outward the gate perpendicular to the gate edge. The $x=0$ is located at the point exactly under the gate edge.}
\label{1}
\end{figure}

Figure~1S shows electron density distribution calculated for several gate bias voltages $U_g = -0.9, -1.2, -2$~V. A pronounced weakening of the CR mode between $U_g=-0.9$~V and $U_g=-1.2$~V corresponds to the full depletion of the near-gate outer region of the 2DES at distances of $0.2$~$\mu$m. At the gate voltage $U_g = -1.2$~V, the density restores to the unperturbed level only at the distance of $x \approx 0.5$~$\mu$m. This distance is comparable to the size of a small region with an extremely strong electric field gradient situated at the circumference of the gate. The size of this inhomogeneity region for our structure geometry is dictated by the distance from the gate to the 2DES, $h=480$~nm~\cite{Mikhailov:06, Mikhailov:11}. This region plays a similar role as the skin layer in the Azbel'-Kaner cyclotron resonance (AK-CR) effect in metals~\cite{Azbel}. Specifically, the gate oscillating electric field in the inhomogeneity region pushes electrons, which circulate the CR orbit in the 2DES ($r_c\approx 1.4$~$\mu$m at the magnetic field of $B=38$~mT), and afterwards, synchronously return back to the region. This mechanism is analogous to what happens in a cyclotron.